\documentclass[12pt]{article}
\usepackage{amsmath}
\usepackage{bm}
\usepackage{color}

\begin{document}
\begin{center}
\begin{large}
{\bf Noncommutative phase space with rotational symmetry and hydrogen atom}
\end{large}
\end{center}

\centerline {Kh. P. Gnatenko \footnote{E-Mail address: khrystyna.gnatenko@gmail.com}, V. M. Tkachuk \footnote{E-Mail address: voltkachuk@gmail.com}}
\medskip
\centerline {\small \it Ivan Franko National University of Lviv, Department for Theoretical Physics,}
\centerline {\small \it 12 Drahomanov St., Lviv, 79005, Ukraine}

\abstract{We construct algebra with noncommutativity of coordinates and noncommutativity of momenta which is rotationally invariant and equivalent to noncommutative algebra of canonical type.
 Influence of noncommutativity on the  energy levels of hydrogen atom is studied in rotationally invariant noncommutative phase space. We find corrections to the levels up to the second order in the parameters of noncommutativity and estimate the upper bounds of these parameters.

}

\section{Introduction}

Studying noncommutativity has obtained recently a great interest because of development of String Theory and Quantum Gravity (see, for instance, \cite{Witten,Doplicher}). Idea of noncommutativity is quite old. The idea has been proposed by Heisenberg and formalized by Snyder in his article published in 1947 \cite{Snyder}.

In recent years physical systems have been intensively studied in the framework of noncommutative classical and quantum mechanics. Among them, for instance,  harmonic oscillator \cite{Hatzinikitas,Kijanka,Jing}, Landau problem \cite{Gamboa1,Horvathy,Dayi,Li,Daszkiewicz1}, gravitational quantum well \cite{Bertolami1,Bastos} classical systems with various potentials \cite{Gamboa,Romero,Mirza,Djemai1,Gnatenko5,Gnatenko10}, many-particle systems \cite{Daszkiewicz,Ho,Djemai,Gnatenko1,Gnatenko2,Daszkiewicz2,Gnatenko9} and many others.

In general noncommutative phase space of canonical type can be realized with the help of the following commutation relations for coordinates and momenta
  \begin{eqnarray}
[X_{i},X_{j}]=i\hbar\theta_{ij},\label{form101}\\{}
[X_{i},P_{j}]=i\hbar(\delta_{ij}+\gamma_{ij}),\label{form1001}\\{}
[P_{i},P_{j}]=i\hbar\eta_{ij},\label{form10001}{}
\end{eqnarray}
where $\theta_{ij}$, $\eta_{ij}$, $\gamma_{ij}$ are elements of constant matrixes, $\theta_{ij}$, $\eta_{ij}$ are called parameters of coordinate and momentum noncommutativity, respectively.

The coordinates $X_i$ and momenta $P_i$  which satisfy (\ref{form101}), (\ref{form10001})  can be represented as
\begin{eqnarray}
X_{i}=x_{i}-\frac{1}{2}\sum_{j}\theta_{ij}{p}_{j},\label{repx}\\
P_{i}=p_{i}+\frac{1}{2}\sum_{j}\eta_{ij}{x}_{j},\label{repp}
\end{eqnarray}
with $x_i$, $p_i$ being coordinates and momenta satisfying the ordinary commutation relations
$[x_{i},x_{j}]=0$,  $[x_{i},p_{j}]=i\hbar\delta_{ij}$, $[p_{i},p_{j}]=0$.
Taking into account (\ref{repx}), (\ref{repp}), one has \cite{Bertolami}
\begin{eqnarray}
[X_{i},P_{j}]=i\hbar\delta_{ij}+i\hbar\sum_k\frac{\theta_{ik}\eta_{jk}}{4}.\label{p1001}{}
\end{eqnarray}
Therefore, parameters $\gamma_{ij}$ are considered to be defined as
  \begin{eqnarray}
\gamma_{ij}=\sum_k\frac{\theta_{ik}\eta_{jk}}{4}.\label{for1}
\end{eqnarray}

It is important to note that noncommutativity of canonical type (\ref{form101})-(\ref{form10001}) causes rotational symmetry breaking \cite{Chaichian,Balachandran1}.
New classes of algebras with noncommutativity of coordinates were proposed to recover the rotational symmetry in noncommutative space. For example, in
\cite{Moreno} the rotational invariance was preserved by foliating
the space with concentric fuzzy spheres.
In \cite{Galikova}  the authors constructed rotationally symmetric noncommutative space  as a
sequence of fuzzy spheres. In the space the exact solution of the
hydrogen atom problem was found.  In \cite{Kupriyanov} the curved
noncommutative space was introduced to maintain the rotational
symmetry and the hydrogen atom spectrum was studied.
In paper \cite{Amorim}  in order to preserve the rotation invariance the authors suggested  promotion of
the parameter of noncommutativity to an operator in Hilbert space and introduced
the canonical conjugate momentum of this
operator.

In our previous paper \cite{Gnatenko6} we considered the idea of generalization of parameter of noncommutativity to a tensor and proposed rotationally invariant algebra with noncommutativity of coordinates.
In the present paper we propose the way to preserve rotational symmetry in a space with noncommutativity of coordinates and noncommutativity of momenta. We construct noncommutative algebra  which is rotationally invariant and equivalent to noncommutative algebra of canonical type (\ref{form101})-(\ref{form10001}). In addition we study energy levels of hydrogen atom in rotationally invariant noncommutative phase space.

Note, that in the framework of noncommutative quantum mechanics the hydrogen atom has been studied in \cite{Ho,Djemai,Bertolami,Chaichian,Chaichian1,Chair,Stern,Zaim2,Adorno,Khodja,Alavi}. In paper
\cite{Chaichian} the authors examined hydrogen atom in noncommutative phase space and found corrections to the energy levels of hydrogen atom  up to the first order in the parameter of coordinate noncommutativity.
In this paper corrections to the Lamb shift within the noncommutative quantum
electrodynamic theory were also obtained. In \cite{Ho} the hydrogen atom was studied
as a two-particle system in a space with noncommutativity of coordinates. The authors studied the case  when particles of opposite charges feel opposite noncommutativity.  In \cite{Chair} the quadratic
Stark effect was examined. New result for shifts
in the spectrum of hydrogen atom in noncommutative space  was
presented in \cite{Stern}.
In \cite{Zaim2} the hydrogen atom
energy levels were calculated in the framework of the
noncommutative Klein-Gordon equation. The
Dirac equation with a Coulomb field was considered in
a  space with noncommutativity of coordinates in \cite{Adorno,Khodja}.
In the paper \cite{Djemai} the authors found corrections to the energy levels of hydrogen atom in noncommutative phase space up to the first order in the parameter of noncommutativity. In \cite{Alavi} the author studied spectrum of Hydrogen atom, Lamb shift and Stark effect in a space with noncommutativity of coordinates and noncommutativity of momenta. Full phase-space noncommutativity in the Dirac equation was considered in \cite{Bertolami}.  In this article the influence of momentum noncommutativity on the spectrum of hydrogen atom was studied.
Hydrogen atom  was also considered in the case of
space-time noncommutativity  in \cite{Balachandran,Stern1,Moumni1}.

 In  contrast to the previous papers in this paper we study hydrogen atom in noncommutative phase space with preserved rotational symmetry.

The article is organized as follows. In Section 2 we consider rotationally invariant algebra corresponding to noncommutative phase space. In Section 3 influence of noncommutativity on the energy levels of hydrogen atom is studied in rotationally invariant noncommutative phase space. Conclusions are presented in Section 4.

\section{Rotationally invariant noncommutative phase space }\label{rozd2}

In order to preserve rotational symmetry in noncommutative phase space let us consider generalization of parameters of noncommutativity $\theta_{ij}$, $\eta_{ij}$ to  tensors. We propose to construct these tensors with the help of additional coordinates $a_i$, $b_i$ and conjugate momenta $p^a_i$, $p^b_i$ of them which correspond to rotationally symmetric systems. In our previous papers we propose to define tensor of coordinate noncommutativity as follows
\begin{eqnarray}
\theta_{ij}=\frac{l_{0}}{\hbar}\varepsilon_{ijk}a_{k}, \label{form130}
\end{eqnarray}
with $l_{0}$ being a constant with the dimension of length \cite{Gnatenko6,GnatenkoConf}.
 For reason of simplicity and  reason of dimension we propose to write tensor of momentum noncommutativity as follows
 \begin{eqnarray}
\eta_{ij}=\frac{p_{0}}{\hbar}\varepsilon_{ijk}p^b_{k},\label{for130}
\end{eqnarray}
 here $p_{0}$ being a constant with the dimension of momentum. Here $a_{i}$ and $p^b_i$  are additional coordinates and additional momenta governed by rotationally symmetric systems. We suppose that these systems are harmonic oscillators with parameters $m_{osc}$ and $\omega$
 \begin{eqnarray}
 H^a_{osc}=\frac{(p^{a})^{2}}{2m_{osc}}+\frac{m_{osc}\omega^{2} a^{2}}{2},\label{form104}\\
 H^b_{osc}=\frac{(p^{b})^{2}}{2m_{osc}}+\frac{m_{osc}\omega^{2} b^{2}}{2}.\label{for104}
 \end{eqnarray}
 We put
 \begin{eqnarray}
\sqrt{\frac{\hbar}{m_{osc}\omega}}=l_{P},\label{form200}
 \end{eqnarray}
 where $l_P$ is the Planck length. We also  consider  the frequency $\omega$ to be very large. This leads to grate distance between energy levels of harmonic oscillators $H^a_{osc}$, $H^b_{osc}$. So, these oscillators put into the ground states remain in them.

From  (\ref{for1}), (\ref{form130}) and (\ref{for130}) we have
  \begin{eqnarray}
\gamma_{ij}=\frac{l_0p_0}{4\hbar^2}\left(({\bf a}\cdot{\bf p^b})\delta_{ij}-a_jp^b_i\right).\label{fom1}
\end{eqnarray}
So, we propose the following noncommutative algebra
\begin{eqnarray}
[X_{i},X_{j}]=i\varepsilon_{ijk} l_{0} a_{k},\label{form131}\\{}
[X_{i},P_{j}]=i\hbar\left(\delta_{ij}+\frac{l_0p_0}{4\hbar^2}({\bf a}\cdot{\bf p^b})\delta_{ij}-\frac{l_0p_0}{4\hbar^2}a_jp^b_i\right),\\{}
[P_{i},P_{j}]=\varepsilon_{ijk} p_{0} p^b_{k}.\label{form13331}{}
\end{eqnarray}
  Additional coordinates $a_{i}$, $b_{i}$ and momenta $p^{a}_{i}$, $p^{b}_{i}$ satisfy the ordinary commutation relations
\begin{eqnarray}
[a_{i},a_{j}]=[b_{i},b_{j}]=[a_{i},b_{j}]=0,{}\\{}
[a_{i},p^{a}_{j}]=[b_{i},p^{b}_{j}]=i\hbar\delta_{ij},{}\\{}
[p^{a}_{i},p^{a}_{j}]=[p^{b}_{i},p^{b}_{j}]=[p^{a}_{i},p^{b}_{j}]=0,{}\\{}
[a_{i},p^{b}_{j}]=[b_{i},p^{a}_{j}]=0.
\end{eqnarray}
  So, $\gamma_{ij}$, $\theta_{ij}$, $\eta_{ij}$, given by  (\ref{fom1}), (\ref{form130}), (\ref{for130}) commute with each other. Also, $a_{i}$ and $p^{b}_i$ commute with $X_{i}$ and $P_{i}$. Therefore,
\begin{eqnarray}
[\theta_{ij}, X_k]=[\theta_{ij}, P_k]=[\eta_{ij}, X_k]=[\eta_{ij}, P_k]=[\gamma_{ij}, X_k]=[\gamma_{ij}, P_k]=0
\end{eqnarray}
So, $X_{i}$, $P_{i}$, $\theta_{ij}$, $\eta_{ij}$ and $\gamma_{ij}$ satisfy the same commutation relations as in the case of the canonical version of noncommutative phase space. In this sense algebra  (\ref{form131})-(\ref{form13331}) is equivalent to algebra (\ref{form101})-(\ref{form10001}). Moreover  algebra (\ref{form131})-(\ref{form13331}) is rotationally invariant. Commutation relations remains the same after rotation
\begin{eqnarray}
[X^\prime_{i},X^\prime_{j}]=i\varepsilon_{ijk} l_{0} a^\prime_{k},\\{}
[X^\prime_{i},P^\prime_{j}]=i\hbar\left(\delta_{ij}+\frac{l_0p_0}{4\hbar^2}({\bf a^\prime}\cdot{\bf p^{b\prime}})\delta_{ij}-\frac{l_0p_0}{4\hbar^2}a^\prime_jp^{b\prime}_i\right),\\{}
[P^{\prime}_{i},P^{\prime}_{j}]=\varepsilon_{ijk} p_{0} p^{b\prime}_{k}.{}
\end{eqnarray}
Here $X_{i}^{\prime}=U(\varphi)X_{i}U^{+}(\varphi)$, $a_{i}^{\prime}=U(\varphi)a_{i}U^{+}(\varphi)$,  $p^{b\prime}_{i}=U(\varphi)p^b_{i}U^{+}(\varphi)$. The rotation operator is the following
\begin{eqnarray}
 U(\varphi)=e^{\frac{i}{\hbar}\varphi({\bf n}\cdot{\bf\tilde{L}})}
 \end{eqnarray}
 with
 \begin{eqnarray}
 {\bf\tilde{L}}=[{\bf r}\times{\bf p}]+[{\bf a}\times{\bf p}^{a}]+[{\bf b}\times{\bf p}^{b}]
 \label{form500}
 \end{eqnarray}
being the total angular momentum. Here ${\bf r}=(x_1,x_2,x_3)$

Making straightforward calculations we can conclude that operator ${\bf\tilde{L}}$ commutes with scalar products
 \begin{eqnarray}
[\tilde{L}_{i},({\bf a}\cdot{\bf p})]=[\tilde{L}_{i},({\bf b}\cdot{\bf p})]=[\tilde{L}_{i},({\bf a}\cdot{\bf b})]=[\tilde{L}_{i},({\bf r}\cdot{\bf a})]=[\tilde{L}_{i},({\bf r}\cdot{\bf b})]=\nonumber\\
=[\tilde{L}_{i},({\bf a}\cdot{\bf L})]=[\tilde{L}_{i},({\bf b}\cdot{\bf L})]=[\tilde{L}_{i},({\bf p^a}\cdot{\bf L})]=[\tilde{L}_{i},({\bf p^b}\cdot{\bf L})]=0.\label{com}
\end{eqnarray}
here $L=[{\bf r}\times{\bf p}]$. Also, ${\bf\tilde{L}}$ commutes with $r^2$, $p^2$, $a^2$, $b^2$, $(p^a)^2$, and $(p^b)^2$. Therefore, the total momentum ${\bf\tilde{L}}$ commutes with the operator of distance $R=\sqrt{\sum_{i}X_{i}^{2}}$ which taking into account (\ref{repx}), (\ref{repp}), (\ref{form130}), (\ref{for130}) can be written in the following form
\begin{eqnarray}
R=\sqrt{\sum_{i}X_{i}^{2}}=\sqrt{r^2+\frac{l_0^2}{4\hbar^2}a^2p^2-\frac{l_0}{4\hbar^2}({\bf a}\cdot{\bf p})^2-\frac{l_0}{\hbar}({\bf a}\cdot{\bf L})}.
\end{eqnarray}
So, the distance remains the same after rotation $R^{\prime}=U(\varphi)R U^{+}(\varphi)=R$.
Also, taking into account (\ref{com}), we have
 \begin{eqnarray}
[\tilde{L}_{i},P]=0
\end{eqnarray}
here
\begin{eqnarray}
P=\sqrt{\sum_{i}P_{i}^{2}}=\sqrt{p^2+\frac{p^2_0}{4\hbar^2}r^2(p^b)^2-\frac{p^2_0}{4\hbar^2}({\bf r}\cdot{\bf p^b})^2+\frac{p_0}{\hbar}({\bf p^b}\cdot{\bf L})}.
\end{eqnarray}
So, the absolute value of momentum remains the same after rotation $P^{\prime}=U(\varphi)P U^{+}(\varphi)=P$.

It is also worth  mentioning that the following commutation relations are satisfied
  \begin{eqnarray}
 [X_{i},\tilde{L}_{j}]=i\hbar\varepsilon_{ijk}X_{k},\\{}
 [P_{i},\tilde{L}_{j}]=i\hbar\varepsilon_{ijk}P_{k},\\{}
  [a_{i},\tilde{L}_{j}]=i\hbar\varepsilon_{ijk}a_{k},\\{}
  [p^{a}_{i},\tilde{L}_{j}]=i\hbar\varepsilon_{ijk}p^{a}_{k},\\{}
  [b_{i},\tilde{L}_{j}]=i\hbar\varepsilon_{ijk}b_{k},\\{}
  [p^{b}_{i},\tilde{L}_{j}]=i\hbar\varepsilon_{ijk}p^{b}_{k},
  \end{eqnarray}
  which are the same as in the ordinary space ($\theta_{ij}=\eta_{ij}=0$).

Explicit representation for noncommutative coordinates $X_i$ and noncommutative momenta $P_i$, which taking into account  (\ref{repx}), (\ref{repp}), (\ref{form130}), (\ref{for130}), reads
\begin{eqnarray}
X_{i}=x_{i}+\frac{l_0}{2\hbar}[{\bf a}\times{\bf p}]_i,\label{repx0}\\
P_{i}=p_{i}-\frac{p_0}{2\hbar}[{\bf r}\times{\bf p^b}]_i,\label{repp0}
\end{eqnarray}
(here coordinates $x_{i}$ and momenta $p_{i}$ satisfy the ordinary commutation relations
and commute with $a_{i}$, $p^a_{i}$) guarantees that the Jacobi identity is satisfied. This can be easily checked for all possible triplets of operators.

Note, that from (\ref{repx0}), (\ref{repp0}) follows that
\begin{eqnarray}
[X_{i},p^a_{j}]=i\varepsilon_{ijk}\frac{l_{0}}{2}p_{k},{}\\{}
[P_{i},b_{j}]=i\varepsilon_{ijk}\frac{l_{0}}{2}x_{k},{}\\{}
[X_{i},a_{j}]=[X_{i},b_{j}]=[X_{i},p^b_{j}]=[P_{i},a_{j}]=[P_{i},p^a_{j}]=[P_{i},p^b_{j}]=0.
\end{eqnarray}

At the end of this section we would like to discuss possible physical meaning of additional coordinates $a_i$, $b_i$ which we considered in order to construct tensors of noncommutativity.
The coordinates can be treated as some
internal coordinates of a particle. Quantum fluctuations of these coordinates lead effectively
to a non-point-like particle, size of which is of the order of the Planck scale.

\section{Hydrogen atom in noncommutative phase space with preserved rotational symmetry}\label{rozd2}

Let us study influence of noncommutativity on the energy levels of hydrogen atom in rotationally invariant noncommutative phase space (\ref{form131})-(\ref{form13331}).
Note that tensors of noncommutativity (\ref{form130}), (\ref{for130}) corresponding to  (\ref{form131})-(\ref{form13331}) are defined with  the help of additional coordinates and momenta which are governed by the harmonic oscillators.
Therefore in order to examine energy levels of hydrogen atom in noncommutative phase space we have to consider the following hamiltonian
 \begin{eqnarray}
H=H_{h}+H^a_{osc}+H^b_{osc},\label{form13}
 \end{eqnarray}
 here
 \begin{eqnarray}
 H_{h}=\frac{P^{2}}{2M}-\frac{e^{2}}{R},
  \end{eqnarray}
where $R=\sqrt{\sum_{i}X_{i}^{2}}$, and $X_i$, $P_i$ satisfy (\ref{form131})-(\ref{form13331}) and $H^a_{osc}$, $H^b_{osc}$ are given by (\ref{form104}), (\ref{for104}).

Using representation (\ref{repx0}), (\ref{repp0}) we can write
 \begin{eqnarray}
 H_{h}=\frac{p^{2}}{2M}+\frac{({\bm \eta}\cdot{\bf L})}{2M}+\frac{[{\bm \eta} \times {\bf r}]^{2}}{8M}-\frac{e^{2}}{\sqrt{r^{2}-({\bm \theta}\cdot{\bf L})+\frac{1}{4}[{\bm \theta}\times{\bf p}]^{2}}},
  \end{eqnarray}
here the following notations
 \begin{eqnarray}
 {\bm \theta}=\frac{l_0}{\hbar}{\bf a},\\
 {\bm \eta}=\frac{p_0}{\hbar}{\bf p^b},
   \end{eqnarray}
   are used for convenience.

Let us find corrections to the energy levels of the hydrogen atom up to the second order in the parameters of noncommutativity. To do that, let us write expansion of the hamiltonian $H_{h}$ in the series over ${\bm \theta}$. Note, that for $1/R$ one has the following expansion
\begin{eqnarray}
\frac{1}{R}=\frac{1}{\sqrt{r^{2}-({\bm \theta}\cdot{\bf L})+\frac{1}{4}[{\bm \theta}\times{\bf p}]^{2}}}=\nonumber\\=\frac{1}{r}+\frac{1}{2r^{3}}({\bm{\theta}}\cdot{\bf L})+\frac{3}{8r^{5}}({\bm{\theta}}\cdot{\bf L})^{2}-
\frac{1}{16}\left(\frac{1}{r^{2}}[{\bm{\theta}}\times{\bf p}]^{2}\frac{1}{r}+\frac{1}{r}[{\bm{\theta}}\times{\bf p}]^{2}\frac{1}{r^{2}}+\frac{\hbar^{2}}{r^{7}}[{\bm{\theta}}\times{\bf r}]^{2}\right).\label{form9}
\end{eqnarray}
Note, that the last term in (\ref{form9}) is caused by noncommutativity of operators $[{\bm \theta}\times{\bf p}]^{2}$ and $r^{2}$ under the square root.
So, using (\ref{form9}) we can write
\begin{eqnarray}
H=H_{0}+V,\label{form41}
\end{eqnarray}
with $V$ being the perturbation caused by the noncommutativity of coordinates
\begin{eqnarray}
V=\frac{({\bm \eta}\cdot{\bf L})}{2M}+\frac{[{\bm \eta} \times {\bf r}]^{2}}{8M}-\nonumber\\-\frac{e^{2}}{2r^{3}}({\bm{\theta}}\cdot{\bf L})-\frac{3e^{2}}{8r^{5}}({\bm{\theta}}\cdot{\bf L})^{2}+
\frac{e^{2}}{16}\left(\frac{1}{r^{2}}[{\bm{\theta}}\times{\bf p}]^{2}\frac{1}{r}+\frac{1}{r}[{\bm {\theta}}\times{\bf p}]^{2}\frac{1}{r^{2}}+\frac{\hbar^{2}}{r^{7}}[{\bm{\theta}}\times{\bf r}]^{2}\right),\label{v}
\end{eqnarray}
and
\begin{eqnarray}
H_{0}=\frac{p^{2}}{2M}-\frac{e^{2}}{r}+H^a_{osc}+H^b_{osc}.\label{form9999}
\end{eqnarray}

Let us find corrections to the energy levels of hydrogen atom caused by noncommutativity.
It is worth mentioning  that
\begin{eqnarray}
\left[\frac{p^{2}}{2M}-\frac{e^{2}}{r},H^a_{osc}\right]=\left[\frac{p^{2}}{2M}-\frac{e^{2}}{r},H^b_{osc}\right]=\left[H^a_{osc},H^b_{osc}\right]=0,
\end{eqnarray}
operator $p^{2}/2M-e^{2}/r$ is the Hamiltonian of the hydrogen atom in the ordinary space ($\theta_{ij}=\eta_{ij}=0$). So, we can write eigenvalues and eigenstates of $H_{0}$
\begin{eqnarray}
E^{(0)}_{n,\{n^{a}\},\{n^{b}\}}=-\frac{e^{2}}{2a_{B}n^{2}}+\hbar\omega(n_{1}^{a}+n_{2}^{a}+n_{3}^{a}+n_{1}^{b}+n_{2}^{b}+n_{3}^{b}+3),\\
\psi^{(0)}_{n,l,m,\{n^{a}\},\{n^{b}\}}=\psi_{n,l,m}\psi^{a}_{n_{1}^{a},n_{2}^{a},n_{3}^{a}}\psi^{b}_{n_{1}^{b},n_{2}^{b},n_{3}^{b}},
\end{eqnarray}
where $\psi_{n,l,m}$, $\psi^{a}_{n_{1}^{a},n_{2}^{a},n_{3}^{a}}$ , $\psi^{b}_{n_{1}^{b},n_{2}^{b},n_{3}^{b}}$  are well known eigenfunctions of the hydrogen atom and the three-dimensional harmonic oscillators $H^a_{osc}$, $H^b_{osc}$ in the ordinary space ($\theta_{ij}=\eta_{ij}=0$),  $a_{B}$ is the Bohr radius.
Let us note ones again that we consider the case when harmonic oscillators $H^a_{osc}$, $H^b_{osc}$ are in the ground states. So, taking this into account and using perturbation theory we can write the corrections to the energy levels of hydrogen atom as follows
\begin{eqnarray}
\Delta E^{(1)}_{n,l}=\langle\psi^{(0)}_{n,l,m,\{0\},\{0\}}|V|\psi^{(0)}_{n,l,m,\{0\},\{0\}}\rangle.
\end{eqnarray}
First, let us mention that
\begin{eqnarray}
\langle\psi^{a}_{0,0,0}|\theta_{i}|\psi^{a}_{0,0,0}\rangle=0,\\
\langle\psi^{b}_{0,0,0}|\eta_{i}|\psi^{b}_{0,0,0}\rangle=0,
\end{eqnarray}
therefore, corrections to the energy levels caused by the terms of the first order in the parameters of noncommutativity   vanish. Namely,
\begin{eqnarray}
\left\langle\psi^{(0)}_{n,l,m,\{0\},\{0\}}\left|\frac{({\bm \eta}\cdot{\bf L})}{2M}\right|\psi^{(0)}_{n,l,m,\{0\},\{0\}}\right\rangle=0,\label{zero1}\\
\left\langle\psi^{(0)}_{n,l,m,\{0\},\{0\}}\left|\frac{e^{2}}{2r^{3}}({\bm{\theta}}\cdot{\bf L})\right|\psi^{(0)}_{n,l,m,\{0\},\{0\}}\right\rangle=0.\label{zero2}
\end{eqnarray}
Let us calculate corrections caused by the term $[{\bm \eta} \times {\bf r}]^{2}/8M$
\begin{eqnarray}
\left\langle\psi^{(0)}_{n,l,m,\{0\},\{0\}}\left|\frac{[{\bm \eta} \times {\bf r}]^{2}}{8M}\right|\psi^{(0)}_{n,l,m,\{0\},\{0\}}\right\rangle=\nonumber\\=\left\langle\psi^{(0)}_{n,l,m,\{0\},\{0\}}\left|\frac{\eta^2 r^2}{8M}-\frac{({\bm \eta}\cdot {\bf r})^{2}}{8M}\right|\psi^{(0)}_{n,l,m,\{0\},\{0\}}\right\rangle=\frac{a_B^2n^2}{24M}(5n^2+1-3l(l+1))\langle\eta^{2}\rangle.\label{eta2}
\end{eqnarray}
Here we take into account that
\begin{eqnarray}
\langle\psi^{b}_{0,0,0}|\eta_{i}\eta_{j}|\psi^{b}_{0,0,0}\rangle=\frac{m_{osc}\omega p_o^2}{2\hbar}\delta_{ij}=\frac{1}{3}\langle\eta^{2}\rangle\delta_{ij},\label{form20}
\end{eqnarray}
with
\begin{eqnarray}
\langle\eta^{2}\rangle=\frac{p_o^2}{\hbar^2}\langle\psi^{b}_{0,0,0}|(p^b)^{2}|\psi^{b}_{0,0,0}\rangle=\frac{3m_{osc}\omega p_o^2}{2\hbar}=\frac{3p_o^2}{2l^2_P}.\label{form887}
\end{eqnarray}
and use the following result (see, for example, \cite{Wen-Chao})
\begin{eqnarray}
\left\langle\psi_{n,l,m}\left|r^2\right|\psi_{n,l,m}\right\rangle=a_B^2\frac{n^2}{2}(5n^2+1-3l(l+1)).
\end{eqnarray}
Also, we have
\begin{eqnarray}
\left\langle\psi^{(0)}_{n,l,m,\{0\},\{0\}}\left|-\frac{3e^{2}}{8r^{5}}({\bm{\theta}}\cdot{\bf L})^{2}+
\frac{e^{2}}{16}\left(\frac{1}{r^{2}}[{\bm{\theta}}\times{\bf p}]^{2}\frac{1}{r}+\frac{1}{r}[{\bm {\theta}}\times{\bf p}]^{2}\frac{1}{r^{2}}+\frac{\hbar^{2}}{r^{7}}[{\bm{\theta}}\times{\bf r}]^{2}\right)\right|\psi^{(0)}_{n,l,m,\{0\},\{0\}}\right\rangle=\nonumber\\=
-\frac{\hbar^{2}e^{2}\langle\theta^{2}\rangle}{a_{B}^{5}n^{5}}\left(\frac{1}{6l(l+1)(2l+1)}-\frac{6n^{2}-2l(l+1)}{3l(l+1)(2l+1)(2l+3)(2l-1)}\right.+\nonumber\\\left.+\frac{5n^{2}-3l(l+1)+1}{2(l+2)(2l+1)(2l+3)(l-1)(2l-1)}-\frac{5}{6}\frac{5n^{2}-3l(l+1)+1}{l(l+1)(l+2)(2l+1)(2l+3)(l-1)(2l-1)}\right),\nonumber\\\label{for411}
\end{eqnarray}
here
\begin{eqnarray}
\langle\theta^{2}\rangle=\frac{l_0^2}{\hbar^2}\langle\psi^{a}_{0,0,0}| a^2|\psi^{a}_{0,0,0}\rangle=\frac{3 l_0^2}{2\hbar}\left(\frac{1}{m_{osc}\omega}\right)=\frac{3 l_0^2l_P^2}{2\hbar^2}.\label{for887}
\end{eqnarray}
The details of calculations of  integrals in (\ref{for411}) can be found in our paper \cite{Gnatenko6}.

So, taking into account (\ref{zero1}), (\ref{zero2}), (\ref{eta2}) and (\ref{for411}) corrections to the energy levels of hydrogen atom in the first order of perturbation theory read
\begin{eqnarray}
\Delta E^{(1)}_{n,l}=\Delta E^{(\eta)}_{n,l}+\Delta E^{(\theta)}_{n,l},\label{form411}
\end{eqnarray}
with
 \begin{eqnarray}
\Delta E^{(\eta)}_{n,l}=\frac{a_B^2n^2\langle\eta^{2}\rangle}{24M}(5n^2+1-3l(l+1)),\label{ft411}
\end{eqnarray}
being corrections caused by noncommutativity of momenta and
\begin{eqnarray}
\Delta E^{(\theta)}_{n,l}=-\frac{\hbar^{2}e^{2}\langle\theta^{2}\rangle}{a_{B}^{5}n^{5}}\left(\frac{1}{6l(l+1)(2l+1)}-\frac{6n^{2}-2l(l+1)}{3l(l+1)(2l+1)(2l+3)(2l-1)}\right.+\nonumber\\\left.+\frac{5n^{2}-3l(l+1)+1}{2(l+2)(2l+1)(2l+3)(l-1)(2l-1)}-\right.\nonumber\\\left.-\frac{5}{6}\frac{5n^{2}-3l(l+1)+1}{l(l+1)(l+2)(2l+1)(2l+3)(l-1)(2l-1)}\right).\label{fe411}
\end{eqnarray}
being corrections caused by coordinates noncommutativity.

In order to find corrections to the energy levels of hydrogen atom up to the second order in the parameters of noncommutativity we need to consider also
the second order of the perturbation theory. We have
\begin{eqnarray}
\Delta
E_{n,l,m,\{0\}}^{(2)}=\sum_{n^{\prime},l^{\prime},m^{\prime},\{n^{a}\},\{n^{b}\}}\frac{\left|\left\langle\psi^{(0)}_{n^{\prime},l^{\prime},m^{\prime},\{n^{a}\},\{n^{b}\}}\left|
V\right|\psi^{(0)}_{n,l,m,\{0\},\{0\}}\right\rangle\right|^{2}}{E^{(0)}_{n}-E^{(0)}_{n^{\prime}}-\hbar\omega(n^{a}_{1}+n^{a}_{2}+n^{a}_{3}+n^{b}_{1}+n^{b}_{2}+n^{b}_{3})},\label{form311}
\end{eqnarray}
here the set of numbers $n^{\prime}$, $l^{\prime}$, $m^{\prime}$,
$\{n^{a}\}$, $\{n^{b}\}$ does not coincide with the set $n$, $l$,
$m$, $\{0\}$, $\{0\}$. We also use notation  $E^{(0)}_{n}$ for the unperturbed energy
of the hydrogen atom
\begin{eqnarray}
E^{(0)}_{n}=-\frac{e^{2}}{2a_{B}n^{2}}.\label{e0}
\end{eqnarray}
The frequency of the harmonic oscillator $\omega$ is considered to be very large. Taking into account that $\left\langle\psi^{(0)}_{n^{\prime},l^{\prime},m^{\prime},\{n^{a}\}}\left| V\right|\psi^{(0)}_{n,l,m,\{0\}}\right\rangle$ does not depend on $\omega$ because of (\ref{form200}). In the  limit $\omega\rightarrow\infty$ we have
\begin{eqnarray}
\lim\limits_{\omega\rightarrow\infty}\Delta
E_{n,l,m,\{0\}}^{(2)}=0.\label{form300}
\end{eqnarray}
Therefore, up to the second order over the parameters of noncommutativity the corrections to the energy levels read
\begin{eqnarray}
\Delta E_{n,l}=\Delta E^{(1)}_{n,l}.\label{form30000}
\end{eqnarray}

It is important to note that obtained result (\ref{form30000}) is divergent in the case when $l=0$ or $l=1$. This means that to find finite result for corrections to $ns$ and $np$ levels we can not use expansion  of hamiltonian into the series over the parameters of noncommutativity. In order to estimate the value of parameters of noncommutativity we are interested in the corrections to the $ns$ energy levels because they are measured with hight precision. To find corrections to the $ns$ energy levels let us rewrite
 perturbation $V$ as follows
  \begin{eqnarray}
 V=\frac{({\bm \eta}\cdot{\bf L})}{2M}+\frac{[{\bm \eta} \times {\bf r}]^{2}}{2M}-\frac{e^{2}}{R}+\frac{e^{2}}{r}=-\frac{e^{2}}{\sqrt{r^{2}-({\bm{\theta}}\cdot{\bf L})+\frac{1}{4}[{\bm{\theta}}\times{\bf p}]^{2}}}+\frac{e^{2}}{r}.\label{form600}
  \end{eqnarray}
Therefore the corrections to these levels read
\begin{eqnarray}
\Delta E_{ns}=\nonumber\\=
\left\langle\psi^{(0)}_{n,0,0,\{0\},\{0\}}\left|\frac{({\bm \eta}\cdot{\bf L})}{2M}+\frac{[{\bm \eta} \times {\bf r}]^{2}}{8M}-\frac{e^{2}}{r}-\frac{e^{2}}{\sqrt{r^{2}-({\bm{\theta}}\cdot{\bf L})+\frac{1}{4}[{\bm{\theta}}\times{\bf p}]^{2}}}\right|\psi^{(0)}_{n,0,0,\{0\},\{0\}}\right\rangle=\nonumber\\=
\frac{a_B^2n^2\langle\eta^{2}\rangle}{24M}(5n^2+1)+1.72\frac{\hbar\langle\theta\rangle\pi e^{2}}{8a_{B}^{3}n^{3}}.\nonumber\\
 \label{form721}
 \end{eqnarray}
 with
 \begin{eqnarray}
\langle\theta\rangle=\langle\psi^{a}_{0,0,0}|\sqrt{\sum_{i}\theta_{i}^{2}}|\psi^{a}_{0,0,0}\rangle=\frac{2l_0 l_{p}}{\sqrt{\pi}\hbar},
\end{eqnarray}
Here we use (\ref{eta2}) and results of calculation of corresponding integrals presented in our papers \cite{GnatenkoKr,GnatenkoConf}.

Analyzing obtained results (\ref{form30000}), (\ref{form721}), we can conclude that there is an important difference between influences of coordinates noncommutativity  and momentum noncommutativity on the energy levels of hydrogen atom. For energy levels with large quantum numbers $n$ we have that corrections caused by noncommutativity of momenta
$\Delta E^{(\eta)}_{n,l}$ (\ref{fe411})
are proportional to $n^4$,  in contrast corrections $\Delta E^{(\theta)}_{n,l}$ (\ref{ft411}) are proportional to $1/n^3$.
So, energy levels with large quantum numbers $n$ are more sensitive to the noncommutativity of momenta than noncommutativity of coordinates. Effect of coordinates noncommutativity better appears for energy levels with small quantum numbers $n$.
Note also that corrections to the $ns$ energy levels (\ref{form721}) include terms  proportional to $\langle\theta\rangle$ and terms proportional to $\langle\eta^{2}\rangle$. In the case of $l>1$ we found that corrections (\ref{form30000}) include terms proportional to $\langle\theta^{2}\rangle$ and  $\langle\eta^{2}\rangle$. So, we can conclude that $ns$ energy levels are more sensitive to the noncommutativity of coordinates (\ref{form131}).

At the end of this section let us estimate the values of parameters of noncommutativity. In order to obtain the upper bounds of the parameters of noncommutativity we suppose that corrections to the hydrogen atom transition energies which are caused by noncommutativity do not exceed the accuracy of the transitions measurements.  Therefore, to estimate upper bounds of the parameters of noncommutativity we use result for $1s-2s$ transition frequency because it is measured with hight precision. In paper  \cite{Parthey} the authors presented   $f_{1s-2s} = 2466061413187018(11)$Hz  with relative uncertainty of $4.5\times10^{-15}$. So, we consider the following inequality
\begin{eqnarray}
\left|\frac{\Delta^{\theta}_{1,2}+\Delta^{\eta}_{1,2}}{E^{(0)}_{2}-E^{(0)}_{1}}\right|\leq4.5\times10^{-15},\label{in}
\end{eqnarray}
here $\Delta^{\theta}_{1,2}$, $\Delta^{\eta}_{1,2}$ are corrections to the energy of $1s-2s$ transition caused by coordinates noncommutativity and momentum noncommutativity, respectively, $E^{(0)}_{n}$ is given by (\ref{e0}).
To estimate the order of values of parameters of coordinate and momentum noncommutativity, it is sufficiently to consider the following inequalities
\begin{eqnarray}
\left|\frac{\Delta^{\theta}_{1,2}}{E^{(0)}_{2}-E^{(0)}_{1}}\right|\leq2.25\times10^{-15},\label{in1}\\
\left|\frac{\Delta^{\eta}_{1,2}}{E^{(0)}_{2}-E^{(0)}_{1}}\right|\leq2.25\times10^{-15}.\label{in2}
\end{eqnarray}

Taking into account (\ref{form721}) we have
\begin{eqnarray}
\Delta^{\theta}_{1,2}=-\frac{3\hbar\langle\theta\rangle\pi e^{2}}{16a_{B}^{3}},\\
\Delta^{\eta}_{1,2}=\frac{13a_B^2\langle\eta^{2}\rangle}{4M}.
\end{eqnarray}

So, we obtain
\begin{eqnarray}
\hbar\langle\theta\rangle\leq10^{-36}\,\textrm{m}^{2},\label{form55}\\
\hbar\sqrt{\langle\eta^{2}\rangle}\leq10^{-61}\,\textrm{kg}^{2}\textrm{m}^{2}/\textrm{s}^{2}.\label{for55}
\end{eqnarray}

It is worth mentioning that our
 estimation for parameters of coordinate and  momentum noncommutativity (\ref{form55}), (\ref{for55}) are in the agreement with the results obtained in the literature. For instance, they are in agreement with upper bounds obtained from the spectrum of gravitation quantum well \cite{Bertolami1}, from the spectrum of hydrogen atom considered in space with noncommutativity of momenta without preserved rotational symmetry \cite{Bertolami}, and on the basis of the data on the Lamb shift
\cite{Chaichian}.

\section{Conclusion}\label{rozd4}

In the paper we have proposed the way to preserve rotational symmetry in a space with noncommutativity of coordinates and noncommutativity of momenta (\ref{form101})-(\ref{form10001}). Rotationally invariant noncommutative algebra was constructed with the help of generalization of parameters of noncommutativity to tensors. The tensors have been defined with the help of additional coordinates and conjugate momenta of them.  The additional coordinates have been considered to be governed by spherically-symmetric systems. For reason of simplicity, we have considered the systems to be harmonic oscillators.  As a result, we have proposed algebra with noncommutativity of coordinates and noncommutativity of momenta (\ref{form131})-(\ref{form13331}) which is rotationally invariant, moreover it is equivalent to the noncommutative algebra of canonical type (\ref{form101})-(\ref{form10001}).

We have studied the hydrogen atom in rotationally invariant noncommutative phase space (\ref{form131})-(\ref{form13331}). We have found corrections to the energy levels of the atom up to the second order in the parameters of noncommutativity (\ref{form30000}).  On the basis of obtained results we conclude that effect of momentum noncommutativity better appears for energy levels of hydrogen atom with large principal quantum numbers.
These levels are more sensitive to the noncommutativity of momenta than noncommutativity of coordinates.
 Effect of coordinates noncommutativity better appears for energy levels with small quantum numbers $n$. In addition we have obtained that corrections to the $ns$-energy levels caused by noncommutativity (\ref{form721}) are proportional to $\langle\theta\rangle$. In the contrast, corrections to energy levels with $l>1$ (\ref{form30000}) are proportional to $\langle\theta^{2}\rangle$.  So, we have concluded that $ns$ energy levels are more sensitive to the noncommutativity of coordinates.

We have estimated the upper bounds of the parameters of noncommutativity on the basis of assumption that corrections to the energies caused by noncommutativity do not exceed the accuracy of  measurements.  So, for this purpose  we have considered hight precision measurement of $1s-2s$ transition of hydrogen atom.
 Comparing corrections to the energy of $1s$-$2s$ transition caused by noncommutativity with the accuracy of experimental results for $1s$-$2s$ transition frequency measurement we have estimated the upper bounds for parameters of coordinate and momentum noncommutativity (\ref{form55}), (\ref{for55}) in rotationally-invariant noncommutative phase space. The upper bounds are in the agreement with that presented in the literature. Among them, for example, are upper bounds obtained from the spectrum of gravitation quantum well \cite{Bertolami1}, from the spectrum of hydrogen atom considered in a space with noncommutativity of momenta without preserved rotational symmetry  \cite{Bertolami}, and on the basis of the data on the Lamb shift
\cite{Chaichian}.

\section*{Acknowledgments}
This work was partly supported by Project FF-30F (No. 0116U001539) from the
Ministry of Education and Science of Ukraine.


\begin{thebibliography}{9}                                                                                                %
\bibitem{Witten} N. Seiberg, E. Witten, J. High Energy Phys. {\bf 9909}, 032 (1999).
\bibitem{Doplicher} S. Doplicher, K. Fredenhagen, J. E. Roberts, Phys. Lett. B {\bf 331}, 39 (1994).


\bibitem{Snyder} H. Snyder, Phys. Rev. {\bf 71}, 38 (1947).
\bibitem{Hatzinikitas} A. Hatzinikitas, I. Smyrnakis, J. Math. Phys. {\bf43},  113 (2002).
\bibitem{Kijanka} A. Kijanka, P. Kosinski, Phys. Rev. D {\bf70},  127702 (2004).
\bibitem{Jing} Jing Jian, Jian-Feng Chen, Eur. Phys. J. C {\bf60}, 669 (2009).



\bibitem{Gamboa1} J. Gamboa, M. Loewe, F. Mendez, J. C. Rojas, Mod. Phys. Lett. A {\bf16}, 2075 (2001).
\bibitem{Horvathy} P. A. Horvathy,  Ann. Phys. {\bf299}  128 (2002).
\bibitem{Dayi} O. F. Dayi, L. T. Kelleyane, Mod. Phys. Lett. A {\bf17}  1937 (2002).
\bibitem{Li} Li Kang, Xiao-Hua Cao, Dong-Yan Wang, Chin. Phys. {\bf15}, 2236 (2006).
\bibitem{Daszkiewicz1} M. Daszkiewicz, Acta Phys. Polon. B {\bf44}, 59 (2013).


\bibitem{Bertolami1} O. Bertolami, J. G. Rosa, C. M. L. de Aragao, P. Castorina, D. Zappala, Phys. Rev. D {\bf72}, 025010 (2005)
\bibitem{Bastos} C. Bastos, O. Bertolami,  Phys. Lett. A {\bf372}, 5556 (2008).

\bibitem{Gamboa} J. Gamboa, M. Loewe, J. C. Rojas, Phys. Rev. D {\bf64},  067901 (2001).
\bibitem{Romero} J.M. Romero, J. D. Vergara, Mod. Phys. Lett. A {\bf18}, 1673 (2003).
\bibitem{Mirza} B. Mirza, M. Dehghani, Commun. Theor. Phys. {\bf42}, 183 (2004).
\bibitem{Djemai1} A. E. F. Djemai, Int. J. Theor. Phys. {\bf43}, 299 (2004).

\bibitem{Gnatenko5} Kh. P. Gnatenko, V. M. Tkachuk, Mod. Phys. Lett. A {\bf 31}, 1650026 (2016).
\bibitem{Gnatenko10} Kh. P. Gnatenko, V. M. Tkachuk, Phys. Lett. A 10.1016/j.physleta.2017.05.056.
\bibitem{Daszkiewicz} M. Daszkiewicz, C. J. Walczyk, Mod. Phys. Lett. A {\bf26}, 819 (2011).

\bibitem{Ho} Pei-Ming Ho, Hsien-Chung Kao, Phys. Rev. Lett. {\bf88}  151602 (2002).
\bibitem{Djemai} A. E. F. Djemai, H. Smail, Commun. Theor. Phys. {\bf41}, 837 (2004).
\bibitem{Gnatenko1} Kh. P. Gnatenko, Phys. Lett. A {\bf 377}, 3061 (2013).
\bibitem{Gnatenko2} Kh. P. Gnatenko, J. Phys. Stud. {\bf17}, 4001 (2013).
\bibitem{Daszkiewicz2} M. Daszkiewicz, Acta Phys. Polon. B {\bf44}, 699 (2013).
\bibitem{Gnatenko9} Kh. P. Gnatenko, V. M. Tkachuk, Ukr. J. Phys. {\bf61},  432 (2016)
\bibitem{Bertolami} O. Bertolami, R. Queiroz, Phys. Lett. A {\bf 375}, 4116 (2011).
\bibitem{Chaichian} M. Chaichian, M. M. Sheikh-Jabbari, A. Tureanu, Phys. Rev. Lett. {\bf 86}, 2716 (2001).
\bibitem{Balachandran1} A. P. Balachandran, P. Padmanabhan,  J. High Energy Phys. {\bf 1012}, 001 (2010).


\bibitem{Moreno} E. F. Moreno, Phys. Rev. D {\bf 72}, 045001 (2005).
\bibitem{Galikova} V. G\'alikov\'a, P. Presnajder, J. Phys: Conf. Ser. {\bf 343}, 012096 (2012).
\bibitem{Kupriyanov} V. G. Kupriyanov, J. Phys. A: Math. Theor. {\bf 46}, 245303 (2013).
\bibitem{Amorim} R. Amorim, Phys. Rev. Lett. {\bf 101}, 081602 (2008).
\bibitem{Gnatenko6} Kh.P. Gnatenko, V. M. Tkachuk, Phys. Lett. A {\bf 378}, 3509 (2014).

\bibitem{Chaichian1} M. Chaichian, M. M. Sheikh-Jabbari, A. Tureanu, Eur. Phys. J. C {\bf 36}, 251 (2004).
\bibitem{Chair} N. Chair, M. A. Dalabeeh, J. Phys. A: Math. Gen. {\bf 38}, 1553 (2005).
\bibitem{Stern} A. Stern, Phys. Rev. Lett. {\bf 100}, 061601 (2008).
\bibitem{Zaim2} S. Zaim, L. Khodja, Y. Delenda, Int. J. Mod. Phys. A  {\bf 26}, 4133 (2011).
\bibitem{Adorno} T. C. Adorno, M. C. Baldiotti, M. Chaichian, D. M. Gitman, A. Tureanu, Phys. Lett.  B {\bf 682}, 235 (2009).
\bibitem{Khodja} L. Khodja, S. Zaim, Int. J. Mod. Phys. A {\bf 27}, 1250100 (2012).



\bibitem{Alavi} S. A. Alavi, Mod. Phys. Lett. A {\bf 22}, 377 (2007).

\bibitem{Balachandran} A. P. Balachandran, A. Pinzul Mod. Phys. Lett. A {\bf 20}, 2023 (2005).
\bibitem{Stern1} A. Stern, Phys. Rev. D {\bf 78}, 065006 (2008).
\bibitem{Moumni1} M. Moumni, A. BenSlama, S. Zaim, J. Geom. and Phys. {\bf61}, 151 (2011).

\bibitem{Wen-Chao} Wen-Chao Qiang, Shi-Hai Dong, Phys. Scripta {\bf 70}, 276 (2004).

\bibitem{GnatenkoKr}  Kh. P. Gnatenko, Yu. S. Krynytskyi, V. M. Tkachuk, Mod. Phys. Lett. A {\bf30}, 1550033  (2015).
\bibitem{GnatenkoConf} Kh. P. Gnatenko, J. Phys.: Conf. Ser. {\bf670}, 012023 (2016).
\bibitem{Parthey} A. Matveev, C. G. Parthey, K. Predehl et al., Phys. Rev. Lett. {\bf 110}, 230801 (2013).
\end{thebibliography}
\end{document}